\documentclass[a4paper,fleqn]{cas-dc}

\usepackage{adjustbox}
\usepackage[numbers]{natbib}
\usepackage{algorithm}
\usepackage{algorithmic}
\usepackage{subcaption}
\usepackage{color}
\usepackage{titlesec}
 \titleformat*{\section}{\color{black}\normalfont\bfseries\large}
 \titlespacing{\subsection}{0pt}{6pt}{4pt}

\begin{document}

\renewcommand{\printorcid}{}

\shorttitle{Neural Network Compression using Adiabatic Quantum Computing}

\shortauthors{Zhehui Wang, Benjamin Chen Ming Choong et~al.}

\title [mode = title]{Is Quantum Optimization Ready? An Effort Towards Neural Network Compression using Adiabatic Quantum Computing}                      
\author[1]{Zhehui Wang}
\fnmark[1]
\author[1]{Benjamin Chen Ming Choong}
\fnmark[1]
\author[2]{Tian Huang}
\author[1]{Daniel Gerlinghoff}
\author[1]{Rick Siow Mong Goh}
\author[3]{Cheng Liu}
\author[1]{Tao Luo}
\cormark[1]
\affiliation[1]{organization={Institute of High Performance Computing (IHPC), Agency for Science, Technology and Research (A*STAR)},
addressline={1 Fusionopolis Way, 16-16 Connexis}, 
                postcode={138632}, 
                country={Singapore}}

\affiliation[2]{organization={Huadian Coal Industry Group Co. Ltd.},country={China}}

\affiliation[3]{organization={Institute of Computing Technology, Chinese Academy of Sciences}, country={China}}                
\fntext[fn1]{Zhehui Wang and Benjamin Chen Ming Choong are co-first authors}
\cortext[cor]{Luo Tao is the corresponding author}
\nonumnote{
This research is supported by the National Research Foundation Singapore, under its Quantum Engineering Programme 2.0 (National Quantum Computing Hub, NRF2021-QEP2-02-P01), and the A*STAR C230917003}

\begin{abstract}
Quantum optimization is the most mature quantum computing technology to date, providing a promising approach towards efficiently solving complex combinatorial problems. Methods such as adiabatic quantum computing (AQC) have been employed in recent years on important optimization problems across various domains. In deep learning, deep neural networks (DNN) have reached immense sizes to support new predictive capabilities. Optimization of large-scale models is critical for sustainable deployment, but becomes increasingly challenging with ever-growing model sizes and complexity. While quantum optimization is suitable for solving complex problems, its application to DNN optimization is not straightforward, requiring thorough reformulation for compatibility with commercially available quantum devices. In this work, we explore the potential of adopting AQC for fine-grained pruning-quantization of convolutional neural networks. We rework established heuristics to formulate model compression as a quadratic unconstrained binary optimization (QUBO) problem, and assess the solution space offered by commercial quantum annealing devices. Through our exploratory efforts of reformulation, we demonstrate that AQC can achieve effective compression of practical DNN models. Experiments demonstrate that adiabatic quantum computing (AQC) not only outperforms classical algorithms like genetic algorithms and reinforcement learning in terms of time efficiency but also excels at identifying global optima.\vspace{-7pt}\end{abstract}

% \begin{highlights}
% \item Research highlights item 1
% \item Research highlights item 2
% \item Research highlights item 3
% \end{highlights}


\maketitle

\section{Introduction}

Quantum computing is an emerging computing technology with the promise of massive parallelism. By acting on a superposition of quantum states, quantum computers navigate and transform high-dimensional spaces simultaneously at unprecedented speed and efficiency. Among algorithms and platforms, quantum optimization is the most mature today, and is closest to commercial viability compared to other quantum technologies. Methods such as adiabatic quantum computing (AQC) using physical qubits for combinatorial optimization have already been realized on available devices such as the D-Wave quantum annealers.

Quantum optimization via AQC can rapidly provide solutions to highly-complex problem spaces. Suppose a problem can be mapped to the interactions of qubits, such that the system energy represents the optimization cost. In this case, AQC yields optimization solutions by reaching the lowest energy state through quantum annealing. Specifically, AQC is well-suited for quadratic unconstrained binary optimization (QUBO) problems~\cite{date2021qubo}, as solution variables can be closely modelled as interacting pairs of qubits. The optimization time scales significantly better with problem size in quantum computing, as it utilizes quantum mechanics to achieve the lowest energy state more efficiently than traditional computing methods~\cite{huang2022benchmarking}.

The scalability advantage of AQC enables quantum optimization to demonstrate high potential as an important future platform in parallel computing. In recent years, AQC has been explored in various applications, including financial portfolio optimization~\cite{grant2021benchmarking}, traffic flow management~\cite{inoue2020traffic}, and warehouse management~\cite{sao2019application}, among others. However, mapping non-QUBO problems to a compatible form for quantum optimization is non-trivial. Several recent works have demonstrated the challenge of reformulating even well-researched problems, such as those in computer vision (CV) ~\cite{birdal2021quantum, gou2021knowledge, zaech2022adiabatic} for suitability with AQC.

In the field of deep learning, the optimization of neural networks is essential for practical and sustainable deployment. The representational capacity of neural network models generally grows by introducing deeper layers and increasing parameter count, enabling large-scale models with hundreds of billions of parameters to achieve state-of-the-art predictive performance. However, these huge models are not only costly to deploy due to their intensive computation requirements, but are typically over-parameterized for their tasks. This observation has motivated many studies on optimizing neural network designs, including size, precision, and structural complexity~\cite{liang2021pruning, han2015deep}.

Recent research has concentrated on fine-grained optimization of neural networks, with model compression~\cite{han2015deep} emerging as one of the most prominent topics. With increasing model size and granularity, fine-grained model compression involves a vast design space that is challenging to navigate. To address this issue, classical approaches such as genetic algorithms~\cite{wang2021evolutionary} or reinforcement learning algorithms~\cite{wang2022edcompress} have been proposed. However, these methods are often time-consuming and can yield sub-optimal results due to local optima. Quantum annealing emerges as a promising technique to solve such problems, owing to its highly efficient computational mechanism and time efficiency.

\begin{figure*}[t]
\resizebox{1.045\linewidth}{!}{\begin{minipage}[t]{1.00\linewidth}
\begin{tabular}{ l l |l c c c c }
\multirow{10}{*}{\adjustbox{lap={\width}{-0.8em}}{\includegraphics[width=0.23\linewidth,page=1]{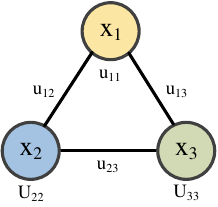}}}& & & \multirow{5}{*}{\raisebox{-0.2\totalheight}{\includegraphics[width=0.12\linewidth,page=1]{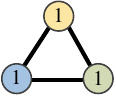}}} & \multirow{5}{*}{\raisebox{-0.2\totalheight}{\includegraphics[width=0.12\linewidth,page=2]{over1.pdf}}} & \multirow{5}{*}{\raisebox{-0.2\totalheight}{\includegraphics[width=0.12\linewidth,page=3]{over1.pdf}}} & \multirow{5}{*}{\raisebox{-0.2\totalheight}{\includegraphics[width=0.12\linewidth,page=4]{over1.pdf}}}\\
  & & &\\
  & & &\\
  & & &\\
  & & &\\
 & &  & $\mathbf{H_1}=u_{11}+u_{22}+u_{33}$ & $\mathbf{H_2} =u_{11}$& $\mathbf{H_3} =u_{22}$& 
 $\mathbf{H_4} =u_{33}$  \\
 & &  & \ \ \ \ $+$  $u_{12}+u_{13}+u_{23}$ & & &    \\
 & & &\multirow{5}{*}{\raisebox{-0.2\totalheight}{\includegraphics[width=0.12\linewidth,page=5]{over1.pdf}}}& \multirow{5}{*}{\raisebox{-0.2\totalheight}{\includegraphics[width=0.12\linewidth,page=6]{over1.pdf}}}& \multirow{5}{*}{\raisebox{-0.2\totalheight}{\includegraphics[width=0.12\linewidth,page=7]{over1.pdf}}}& \multirow{5}{*}{\raisebox{-0.2\totalheight}{\includegraphics[width=0.12\linewidth,page=8]{over1.pdf}}}\\
  & & &\\
  &  & &\\
  & && \\
 $\mathbf{H}= u_{11}x_1^2 + u_{22}x_2^2 + u_{33}x_3^2$ & && \\
  $+u_{12}x_1x_2+u_{13}x_1x_3+u_{23}x_2x_3$  & & & $\mathbf{H_5}= u_{11}+u_{22}+u_{12}$ &$\mathbf{H_6}=u_{11}+u_{33}+u_{13}$&$\mathbf{H_7}= u_{22}+u_{33}+u_{23}$&$\mathbf{H_8}= 0$\\
\vspace{-3pt}\\
\multicolumn{1}{c}{(a)}& \multicolumn{2}{c}{ }  & \multicolumn{4}{c}{(b)}\\
\end{tabular}
\end{minipage}}
\caption{ (a) An example coupling map of three qubits. The cost function (energy) of the system, $\textbf{H}$, can be formulated as a QUBO problem. (b) Since each qubit has two possibilities, the system can exhibit eight possible values of $\textbf{H}$. The AQC paradigm uses quantum mechanics to evolve qubit configurations toward the global minimum energy, seeking the minimum value of $\textbf{H}$.}
\label{f:overview}
\end{figure*}

However, the problem formulations in deep learning do not align directly with quantum optimization techniques, though AQC exhibits potential in handling complex optimizations. In this study, we assess the capabilities of quantum optimization in deep learning by examining its effectiveness in fine-grained neural network compression. Overall, the contributions of this work can be summarized as follows.

\vspace{-0.5em}

\begin{itemize}
    \setlength\itemsep{-0.2em}
    
\item We reformulate the model compression problem to strictly conform to the QUBO structure, ensuring variables are quadratically linked and unconstrained, a key factor for enhancing AQC solution quality.

\item  We evaluate the efficacy of quantum-based model compression algorithms on state-of-the-art annealing hardware, demonstrating that adiabatic quantum computing (AQC) is effective in real-world applications.

\item  We discuss the future of quantum computing in machine learning, emphasizing critical challenges that need addressing to enable broader adoption of quantum optimization techniques. 
\end{itemize}

\section{Related Work}

Quantum optimization has potential applications in a variety of fields, including graph partitioning problems~\cite{ushijima2017graph}, max-cut problem~\cite{poljak1995maximum}, and max-sat problem~\cite{bian2017solving}, and is especially popular among the quantum computing research community because of the simplicity of the problem settings. Applications with more advanced problem settings such as portfolio optimization~\cite{grant2021benchmarking}, traffic flow management~\cite{inoue2020traffic}, and warehouse management problem~\cite{sao2019application} have recently attracted more attention. Furthermore, the reformulation of problems to QUBO for computer vision applications has been demonstrated in recent works to be non-trivial~\cite{birdal2021quantum, zaech2022adiabatic}. Hence, our work aims to make progress towards adapting quantum optimization technologies with it promise of high parallelism to the ongoing and important research area of model compression. 

Fine-grained neural network compression, including joint pruning-quantization, has drawn more attention in recent years. CLIP-Q~\cite{tung2018clip} is among the first works to perform pruning and quantization jointly. Moreover, pruning-quantization was performed during model fine-tuning, dynamically adapting to changes during the training process. The authors of~\cite{yang2020automatic} presented ANNC as a framework to automatically reach a target average compression rate without the manual setting of hyperparameters. Lastly, OPQ~\cite{hu2021opq} proposed an analytical approach to determine a final pruning and quantization solution before any fine-tuning is performed, eliminating repeated compression optimization during fine-tuning. To the best of our knowledge, the joint pruning-quantization of large-scale models has yet to be studied given the high resource intensity required.

\section{Background}

Quantum optimization entails the use of quantum computing technologies to solve optimization problems. A particular formulation that is suitable for quantum optimization is known as the quadratic unconstrained binary optimization (QUBO) problem~\cite{date2021qubo}, in which we solve problems of the form denoted in Equation~(\ref{eq:qubo}):
\begin{equation} \label{eq:qubo}
    \min_{\mathbf{x} \in \{0,1\}^n} \mathbf{x}^T \mathbf{U}\mathbf{x}
\end{equation}
where $\mathbf{x}$ is a binary vector of solution variables of length $n$, and $\mathbf{U}$ is an $n \times n$ unitary matrix encoding the cost function (energy) of the optimization problem. The QUBO form is equivalent to the Ising model of the interaction of atomic spins. In the literature it has 
\textcolor{blue}{been} shown that solving a QUBO problem is NP-hard~\cite{barahona1982computational}.

\subsection{Adiabatic Quantum Computing}

\begin{figure}[t]
  \centering
\includegraphics[width=1\linewidth] {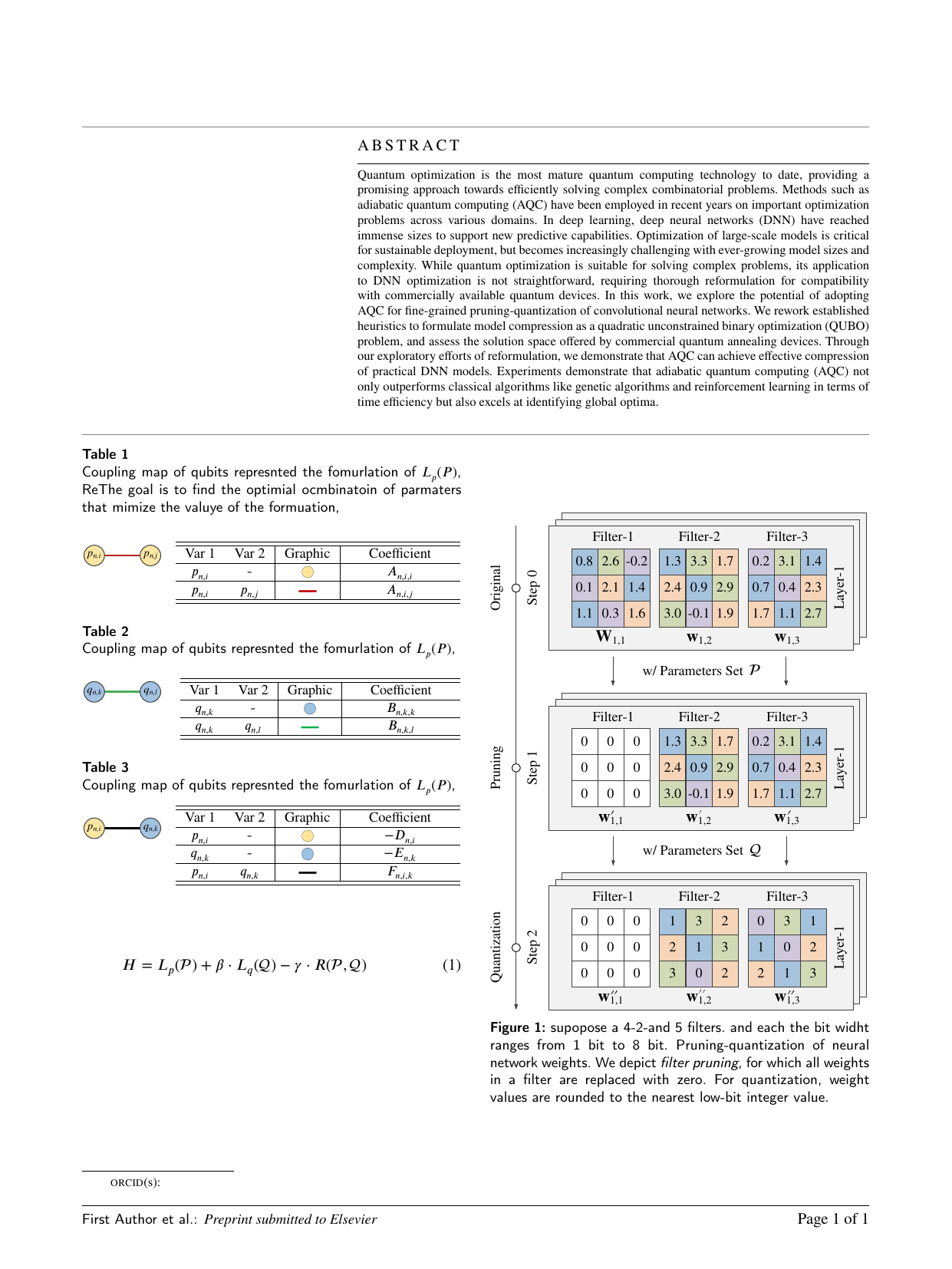} 
  \caption{Joint pruning and quantization of neural network weights. For filter pruning, weights in some filters are replaced with zero. For quantization, weight values are rounded to the nearest low-bit integer value. The challenge is to find the optimal parameter sets $\mathcal{P}$ and $\mathcal{Q}$ for pruning and quantization.}
  \label{f:nn-compression}
\end{figure}

Adiabatic quantum computing (AQC) is a specialized quantum optimization technique that relies on quantum mechanics to solve QUBO problems. The QUBO problem can be seen as a graph of weighted binary nodes connected by weighted edges. Figure~\ref{f:overview}(a) illustrates an example QUBO graph with binary nodes ${x_{i}}$ and node/edge weights $u_{ij}$. A problem formulated as a QUBO can be implemented on quantum hardware by taking the QUBO graph as the coupling map for qubits. The nodes and weights of the QUBO graph correspond to the qubits and couplers in the quantum hardware. While each binary node in the graph represents a \emph{logical qubit}, the true number of \emph{physical qubits} needed for implementation may increase due to constraints in qubit connectivity on the quantum device. 

As each qubit can take on two possible values, the example coupling map of qubits can exhibit eight possible values of $H$, as shown in Figure~\ref{f:overview}(b). The AQC paradigm utilizes quantum mechanics to evolve the qubit configurations towards the global minimum energy, searching for the minimum energy, i.e., the minimal value of $H$. In operation, a quantum Hamiltonian $H_P$, of which the ground state encodes the solution to a problem of interest, is evolved from another Hamiltonian $H_0$, of which the ground state can be easily prepared. By adiabatically evolving the Hamiltonian from $H_0$ to $H_P$, the quantum system will remain in the ground state across the evolution time, allowing the solution of the problem to be reached. 

The specialized quantum hardware used for AQC is known as a quantum annealer. One of the most well-known AQC devices is the D-Wave system~\cite{dwave_arch}, which is based on superconducting loops. While the D-Wave devices use quantum interactions of physical qubits, their qubit numbers are currently limited. In order to simulate larger-scale AQC in the meantime, CMOS-based quantum-inspired annealers are used.  For example, Digital Annealer (DA)~\cite{aramon2019physics} is a specialized computing technology developed by Fujitsu that utilizes quantum-inspired algorithms to solve combinatorial optimization problems. We use Fujitsu's device when problem sizes exceed the current capacity of available D-Wave devices.

\subsection{Neural Network Compression}

Neural network compression refers to the process of reducing the size of a neural network without significantly affecting its performance. Smaller models require less memory and computation resources, making them more suitable for deployment on devices with limited resources, such as mobile phones or Internet of Things (IoT) devices. There are various techniques for compressing neural networks. \emph{Pruning}~\cite{liang2021pruning} removes the least important connections between neurons. Structured pruning considers parameters in coarser-grained groups, removing entire filters or channels to exploit hardware and software optimized for dense computation. \emph{Quantization}~\cite{liang2021pruning} reduces the precision of weights and activations in the network. Lastly, \emph{knowledge distillation}~\cite{gou2021knowledge} trains a smaller student network to mimic the behavior of a larger teacher network.

Figure~\ref{f:nn-compression} provides an illustration of the {processes in model compression}. In this work, we focus on pruning and quantization, both of which are widely accepted as an important step of efficient neural network deployment today. As illustrated in the figure, identifying the optimal parameter set $\mathcal{P}$ and $\mathcal{Q}$ for both pruning and quantization to achieve optimal solutions remains an unresolved problem. While pruning and quantization are well-researched individually, combining these two compression methods increases the complexity of the design space, making the joint pruning-quantization of large-scale models challenging. By reformulating the optimization problem into a QUBO format, we aim to explore the potential of quantum annealing to address this challenge in the future.

\section{Mapping to Quantum Bits}

Before performing quantum optimization, we need to map the parameters of model compression into quantum bits (qubits), which support only binary choices. To achieve this, we formulate the task of joint pruning and quantization. We assume that the model's compression operations are applied to a pre-trained CNN with multiple layers. As we perform joint pruning and quantization, the entire model is fine-tuned once the parameters in both sets $\mathcal{P}$ and $\mathcal{Q}$ are determined

{We develop a method to represent the pruning and quantization parameters using discrete binary representations. For pruning, we employ a binary vector to represent the pruning status of the weight matrix. Each bit of this binary vector indicates whether the corresponding filter in the weight matrix should be pruned. For quantization, we use a binary code to represent the number of bits to be removed from the original data bit-width.}

\begin{figure}[t]
  \centering
  \includegraphics[width=1\linewidth] {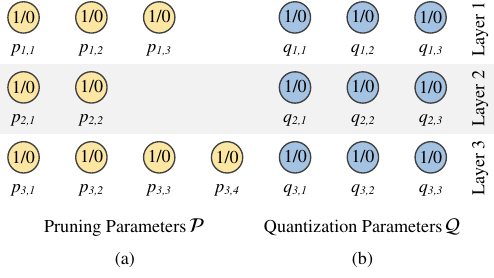} 
  \caption{(a) Qubits in yellow are mapped from the pruning parameters set $\mathcal{P}$, and (b) qubits in blue are mapped from the quantization parameters set $\mathcal{Q}$. In this example, we assume a model with three layers, containing $3$, $2$, and $4$ filters respectively, with weights represented by 8 bits.}
  \label{f:qubits}p
\end{figure}

\subsection{Mapping of Pruning Parameters}

Pruning is the method of selectively removing weights, effectively eliminating redundant connections within neural networks. We establish a collection of pruning masks, which are then applied to the model at a consistent granularity level, such as per weight, channel, or filter, within the weight tensor. Formally, consider the scenario of filter pruning, where the $i$-th filters in the $n$-th layer are targeted. The pruning operation is detailed in Equation~(\ref{eq:prune}).
\begin{equation} \label{eq:prune}
     \mathbf{w}^{\prime}_{n,i}  = (1 - p_{n,i} ) \cdot \mathbf{w}_{n,i} 
\end{equation}
Here, $\textbf{w}_{n,i}$ represents the group of weights in filter $i$ of layer $n$, while $p_{n,i}$ denotes the corresponding mask applied to that filter. Each parameter $p_{n,i}$ can take a binary value, either $0$ or $1$. Specifically, $p_{n,i} = 1$ indicates that the associated filter is pruned, while $p_{n,i} = 0$ signifies that the filter is retained.

With increasing pruning granularity (filter $\rightarrow$ channel $\rightarrow$ weight), the number of design variables $p_{n,i}$ in a typical neural network layer greatly increases.
In our experiments, we find that for deep models, current quantum annealers can only support more coarse-grained \emph{filter pruning} due to constraints on qubit numbers and connectivity.
We apply more fine-grained \emph{channel pruning} only to smaller models that fit within these constraints. Pruning of larger models at a finer granularity would become possible as available qubit numbers increase in the time to come. 

We can map the mask parameters to qubits. Let $\mathcal{P}$ denote the set of pruning variables, as defined in Equation~(\ref{eq:mp}), where each parameter $p_{n,i}$ corresponds to the qubits employed in creating a mask for filter $i$ in layer $n$. Figure~\ref{f:qubits}(a) shows an example of qubits mapped from the pruning parameters set $\mathcal{P}$.
\begin{equation} \label{eq:mp}
\mathcal{P} = \bigcup_{n}\bigcup_{i} \{p_{n,i}\}, \ \ \ \ \ p_{n,i}\in \{ 0, 1\}
\end{equation}

\subsection{Mapping of Quantization Parameters}

Quantization is the process of reducing the precision of weights to decrease the storage footprint. For a specified bit-width, we apply the learned step size quantization (LSQ) method~\cite{esser2019learned} to the convolution weights of the model. This method allows for the learning of the quantization step-size, which is the magnitude between each quantized level. The quantization operation is detailed in Equation~(\ref{eq:lsq}). 
\begin{equation} \label{eq:lsq}
\textbf{w}_{ni}^{\prime \prime} =  \textrm{clamp} \big( \lfloor \frac{\textbf{w}^{\prime}_{n,i}} {s_{n}} \rceil , \, - 2^{b_n}, \,  2^{b_n}-1 \big) \cdot s_{n}
\end{equation}
Here, $\textbf{w}^{\prime}_{n,i}$ represents the weight in the $i$-th filter of the $n$-th layer in the model, following the pruning operations detailed in Equation~(\ref{eq:prune}). Parameter $b_n$ denotes the bit-width of the quantized weight in the $n$-th layer. $s_{n}$ is the learned step-size for layer $n$. The operator $\bigl \lfloor x \bigr \rceil$ rounds $x$ to the nearest integer. The function $\textrm{clamp}(x, u_{0}, u_{1})$ restricts values below $u_{0}$ to $u_{0}$ and those above $u_{1}$ to $u_{1}$. During the optimization process, we fixed $s_n$ for 8-bit quantization. For training and fine-tuning, the step-size for each layer is learned through back-propagation~\cite{esser2019learned}. %When a new precision (bit-width $b_n$) is assigned to a layer during optimization, a new step-size $s_n$ is determined to minimize the mean-squared error (MSE) of the weights resulting from quantization. In real application, this new step-size can be quickly recalculated using equations provided in .

As the search for each $b_{n}$ involves assigning integer values (non-binary), an additional encoding step is necessary to represent $b_n$. This process can be expressed in Equation~(\ref{eq:deb}).
\begin{equation} \label{eq:deb}
b_n = b_{max}- \sum_k q_{n,k} \cdot 2^k
\end{equation}
Here, $b_{max}$ is the original bit-width of the weight before quantization. The binary coded data $(q_{n,k} \dots q_{n,1} q_{n,0})_2$ represents a non-negative integer, calculated as $(q_{n,k} \cdot 2^k + \dots + q_{n,1} \cdot 2^1 + q_{n,0} \cdot 2^0)$. We define this value as the number of bits \emph{removed} from the initial precision due to quantization. As an example, if $b_{max}$ equals eight and $b_{n}$ equals three, we obtain $ (q_{n,2}\,q_{n,1}\,q_{n,0})_{2} = (101)_{2}$ representing five bits removed from the original full precision $b_{max}$.

We can map the binary bits representing $b_n$ to qubits . Let $\mathcal{Q}$ denote the set of quantization variables, as defined in Equation~(\ref{eq:mq}), where each parameter $q_{n,k}$ corresponds to the qubits used to encode quantization bits $b_n$ for layer $n$. Figure~\ref{f:qubits}(b) shows an example of qubits mapped from the quantization parameters set $\mathcal{Q}$.
\begin{equation}\label{eq:mq}
\mathcal{Q} = \bigcup_{n}\bigcup_{k} \{q_{n,k}\}, \ \ \ \ \ q_{n,k}\in \{ 0, 1\}
\end{equation}

\renewcommand{\arraystretch}{1.5}

\begin{table}[h]
\caption{Coupling map of qubits representing the formulation of $L_p(\mathcal{P})$. The goal is to find the optimal combination of parameters that minimize the value of $L_p(\mathcal{P})$} \label{table:qubo1}
\vspace{5pt}
\begin{minipage}{0.236\linewidth}
\centering
\resizebox{1.00\linewidth}{!}{\includegraphics[scale=1]{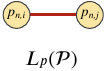}
}
\end{minipage}
\begin{minipage}{0.745\linewidth}
\resizebox{1.00\linewidth}{!}{\begin{tabular}{c c|c|c}
\hline
\hline
{\textrm{Var 1}}&{\textrm{Var 2}}  & {\textrm{Graphic}} & {\ \ \ \ \textrm{Coefficient}\ \ \ \  }  \\
\hline
$p_{n,i}$ & - & \raisebox{-0.2\totalheight}{\includegraphics[width=4mm]{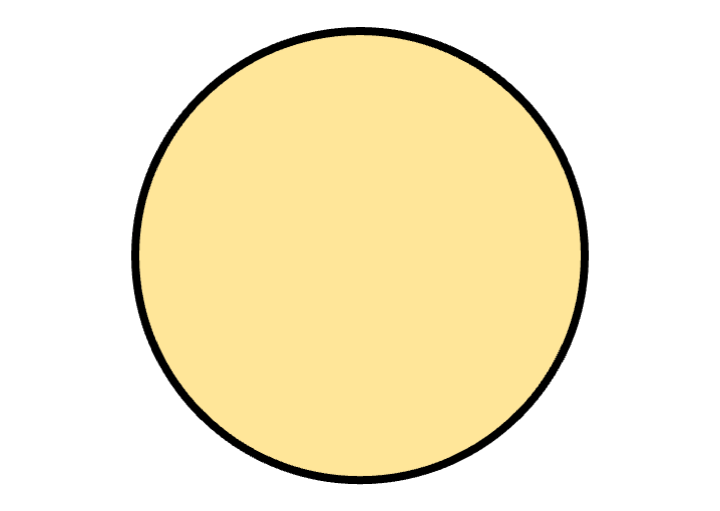}} & $ A_{n,i,i}$  \\
\hline
$p_{n,i}$ & $p_{n,j}$ & \raisebox{-0.3\totalheight}{\includegraphics[width=4mm]{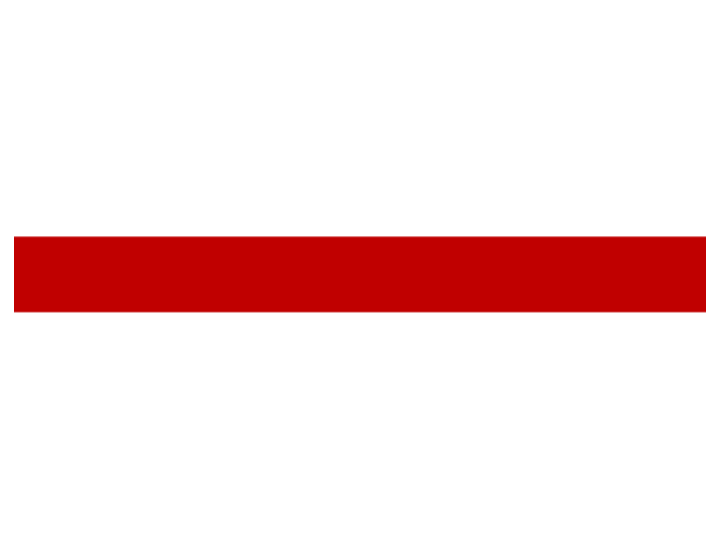}} & $A_{n,i,j}$ \\
\hline
\hline
\end{tabular}}
\end{minipage}
\caption{Coupling map of qubits representing the formulation of $L_q(\mathcal{Q})$}
\label{table:qubo2}
\begin{minipage}{0.236\linewidth}
\centering
\resizebox{1.00\linewidth}{!}{\includegraphics[scale=1]{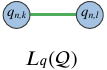}
}
\end{minipage}
\begin{minipage}{0.745\linewidth}
\resizebox{1.00\linewidth}{!}{
\begin{tabular}{c c|c|c}
\hline
\hline
{\textrm{Var 1}}&{\textrm{Var 2}}  & {\textrm{Graphic}} & {\ \ \ \ \textrm{Coefficient}\ \ \ \  }  \\
\hline
$q_{n,k}$ & - & \raisebox{-0.2\totalheight}{\includegraphics[width=4mm]{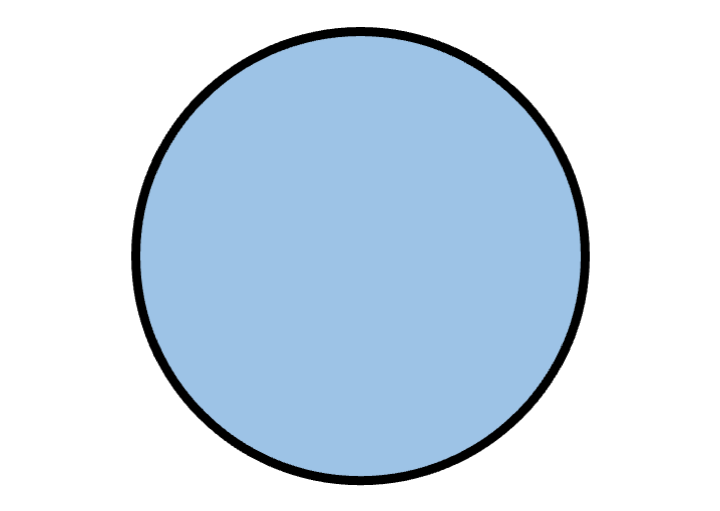}} & $ B_{n,k,k}$  \\
\hline
$q_{n,k}$ & $q_{n,l}$ & \raisebox{-0.3\totalheight}{\includegraphics[width=4mm]{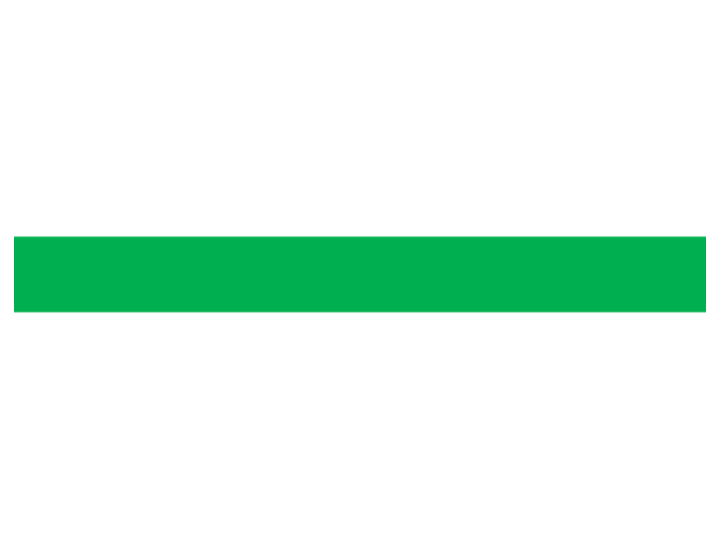}} & $B_{n,k,l}$ \\
\hline
\hline
\end{tabular}}
\end{minipage}
\caption{Coupling map of qubits representing the formulation of $R(\mathcal{P},\mathcal{Q})$} \label{table:qubo3}
\begin{minipage}{0.236\linewidth}
\centering
\resizebox{1.00\linewidth}{!}{\includegraphics[scale=1]{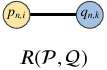}
}
\end{minipage}
\begin{minipage}{0.745\linewidth}
\resizebox{1.00\linewidth}{!}{\begin{tabular}{c c|c| c}
\hline
\hline
{\textrm{Var 1}}&{\textrm{Var 2}}  & {\textrm{Graphic}} & {\ \ \ \ \textrm{Coefficient}\ \ \ \  }  \\
\hline
$p_{n,i}$ & - & \raisebox{-0.2\totalheight}{\includegraphics[width=4mm]{qubo_in_table_1.png}} & $ D_{n,i}$  \\
\hline
$q_{n,k}$ & - & \raisebox{-0.2\totalheight}{\includegraphics[width=4mm]{qubo_in_table_2.png}} & $ E_{n,k}$  \\
\hline
$p_{n,i}$ & $q_{n,k}$ & \raisebox{-0.3\totalheight}{\includegraphics[width=4mm]{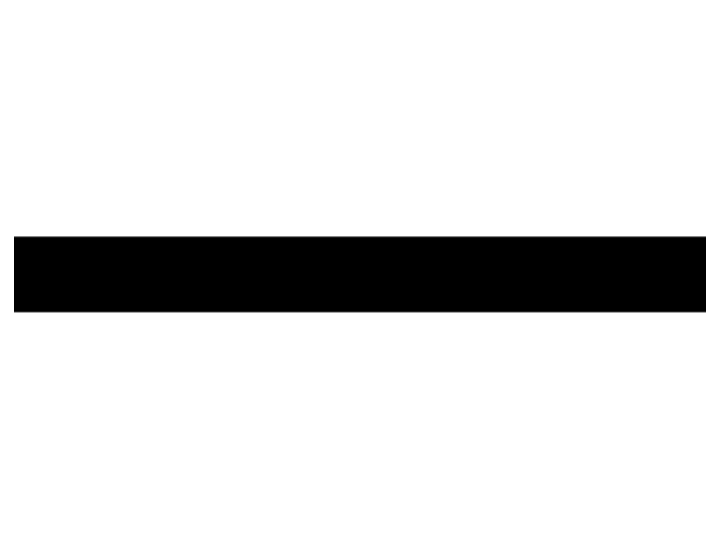}} & $-F_{n,i,k}$ \\
\hline
\hline
\end{tabular}}
\end{minipage}
\end{table}

\section{QUBO Formulation}

The aim of joint pruning-quantization is to determine the value of parameters in the pruning set $\mathcal{P}$ and the quantization set $\mathcal{Q}$ that satisfy design constraints. Ideally, the compression algorithm should provide a design knob for users to trade off accuracy and compression. However, the complete solution space is given by $2^{|\mathcal{P}|} \cdot 2^{|\mathcal{Q}|}$,  growing exponentially with the number of layers and/or granularity of compression (lengths of $\mathcal{P}$ and $\mathcal{Q}$). This challenge has prompted us to reformulate the problem, enabling the use of adiabatic quantum computing to perform the compression task, as quantum technology is particularly adept at solving such problems.

Quantum annealing is an energy minimization process. In the previous section, we encoded the parameters within the pruning sets $\mathcal{P}$ and $\mathcal{Q}$ as interacting binary variables, mapping each variable to a logical qubit in the annealer. In this section, we will formulate our optimization target as the cost function associated with these qubits. Upon measuring the annealing result, each qubit collapses to a binary state, yielding the parameters for pruning and quantization that minimizes the cost function. In the subsequent subsections, we will also elaborate on the coupling map of qubits on the annealing devices 

For our objective, we define three energy components: (i) accuracy loss due to pruning error, (ii) accuracy loss due to quantization error, and (iii) model reduction rate, which is the percentage of weight bits reduced from the original size. The two components of loss add energy to the system, whereas the third compression component removes energy to promote a higher degree of bits removed. Conceptually, our approach is to minimize the cost function (Hamiltonian) $H$ as described in Equation~(\ref{eq:hamiltonian-abstracted}), where $L_{p}$, $L_{q}$ and $R$ are the pruning loss, quantization loss and model reduction ratio respectively, all as functions of qubits in $\mathcal{P}$ and $\mathcal{Q}$.
\begin{equation} \label{eq:hamiltonian-abstracted}
    H = L_{p} ( \mathcal{P} ) + \beta \cdot L_{q} ( \mathcal{Q} ) - \gamma \cdot R ( \mathcal{P} , \mathcal{Q}) 
\end{equation}

We incorporate soft constraints in the optimization process by adding penalties to the objective function. In Equation~(\ref{eq:hamiltonian-abstracted}), the third term, $R ( \mathcal{P} , \mathcal{Q})$, motivates the optimizer to target a higher compression rate.
The first two terms, $L_{p} ( \mathcal{P} ) $ and $L_{q} ( \mathcal{Q} ) $, act as penalties. As the quantization bits decrease, $L_{p} ( \mathcal{P} ) $ and $L_{q} ( \mathcal{Q} ) $ increase, moving in the opposite direction to the objective. This mechanism ensures that the quantization bits do not become too small, thus helping to control the model's accuracy drop.

\begin{table}[h]
\caption{Coupling map of qubits represents the formation of $H$, which is a linear combination of coupler maps for $L_p(\mathcal{P})$, $L_q(\mathcal{Q})$, and $R(\mathcal{P,Q})$. The goal is to find the optimal combination of parameters that minimize the formations for $H$} \label{table:qubo}
\vspace{5pt}
\begin{minipage}{0.236\linewidth}
\centering
\resizebox{1.00\linewidth}{!}{\includegraphics[scale=1]{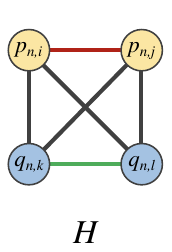}
\label{fig:pies}}
\end{minipage}
\begin{minipage}{0.745\linewidth}
\resizebox{1.00\linewidth}{!}{\begin{tabular}{c c|c|l}
\hline
\hline
{\textrm{Var 1}}&{\textrm{Var 2}}  & {\textrm{Graphic}} & {\ \ \ \ \textrm{Coefficient}\ \ \ \  }  \\
%\cline{1-2}
\hline
$p_{n,i}$ & - & \raisebox{-0.2\totalheight}{\includegraphics[width=4mm]{qubo_in_table_1.png}} & $ A_{n,i,i}-\gamma \cdot {D}_{n,i}$  \\
\hline
$q_{n,k}$ & - & \raisebox{-0.2\totalheight}{\includegraphics[width=4mm]{qubo_in_table_2.png}} &  $\beta \cdot B_{n,k,k}-\gamma \cdot {E}_{n,k}$\\
\hline
$p_{n,i}$ & $p_{n,j}$ & \raisebox{-0.3\totalheight}{\includegraphics[width=4mm]{qubo_in_table_3.png}} & $A_{n,i,j}$ \\
\hline
$q_{n,k}$ & $q_{n,l}$ & \raisebox{-0.3\totalheight}{\includegraphics[width=4mm]{qubo_in_table_5.png}} &  $\beta \cdot {B}_{n,k,l}$ \\
\hline
$p_{n,i}$ & $q_{n,k}$ & \raisebox{-0.3\totalheight}{\includegraphics[width=4mm]{qubo_in_table_4.png}} &  $\gamma \cdot {F}_{n,i,k}$ \\
\hline
\hline
\end{tabular}}
\end{minipage}
\end{table}

\subsection{Pruning Loss}

Following the magnitude-based pruning criterion~\cite{zhu2017prune}, we use the weight magnitude (WM) to approximate the relative accuracy loss resulting from the pruning of a group of weights. The WM can be expressed in Equation~(\ref{eq:pruning-wm}).
\begin{align} \label{eq:pruning-wm}
\text{WM}_n = \sum_i\frac{\|\mathbf{w}_{n,i}\|_{1}}{N_{n,i}} p_{n,i} 
\end{align}
Here, ${{\|\mathbf{w}_{n,i}\|}_1}/{N_{n,i}}$ represent the average magnitude of weights in a filter. The operator ${\|\cdot\|}_1$ is the $l_1$-norm, and $N_{n,i}$ is the number of elements in the weight group $\mathbf{w}_{n,i}$. We assume that when the WM is increased, i.e., the pruned weight magnitude is greater, the accuracy loss due to pruning these weights would be larger.

To convert the formulation of WM into QUBO form, we square the WM to obtain the pruning loss component, i.e., $L_{p}(\mathcal{P})$, which is expressed in Equation (\ref{eq:pruning-loss}). The coefficients of two qubits $p_{n,i}$ and $p_{n,j}$ are represented by $A_{n, i, j}$ and can be pre-calculated before conducting quantum optimization. Accordingly, the coupling map of qubits is shown in Table~\ref{table:qubo1}, which facilitates the implementation of the QUBO formulation into annealing machines.
\begin{align} \label{eq:pruning-loss}
   L_{p} ( \mathcal{P} ) &= \sum_n \text{WM}_n^2 =\sum_{n} \sum_{i} \sum_{j} A_{n, i, j} \cdot \underline{p_{n, i}} \cdot \underline{p_{n, j}}  
\end{align}
\vspace{-15 px}
\begin{align} 
\text{where}\ A_{n, i, j} = \frac{\|\mathbf{w}_{n,i}\|_{1}  \|\mathbf{w}_{n,j}\|_{1} } {N_{n,i}  N_{n,j}} \nonumber
\end{align}

\subsection{Quantization Loss}

We use local quantization error as a proxy for relative accuracy loss due to quantization. We refer to the root mean square error (RMSE) of weights as the indicator of quantization error. The RMSE can be expressed in Equation~(\ref{eq:RMSE}).
\begin{align} \label{eq:RMSE}
\text{RMSE}_n \approx s_n/\sqrt{12}\approx c_n\cdot 2^{\sum_k 2^k q_{n,k}}
\end{align} 
Here, $s_n$ is the step size. It is approximately proportional to $2^{\sum_k 2^k q_{n,k}}$. We use the coefficient $c_n$, a constant for layer $n$, to denote the ratio.

To convert the formulation of RMSE into QUBO form, we take the logarithm of RMSE$/c_n$ and square the result to obtain the quantization loss component, i.e., $L_{q}(\mathcal{Q})$, which is expressed in Equation~(\ref{eq:quantization-loss}). The coefficients of two qubits, $q_{n,k}$ and $q_{n,l}$, are represented by $B_{n,k,l}$ and can be pre-calculated before decoding quantum optimization. Accordingly, the coupling map of qubits is shown in Table~\ref{table:qubo2}.
\begin{align} \label{eq:quantization-loss}
L_{q} ( \mathcal{Q} ) = &\sum_n \log_2^2 (\text{RMSE}_n/c_n) \nonumber \\
        \approx & \sum_{n} \sum_{k} \sum_{l} \underline{B_{n, k, l}} \cdot q_{n, k} \cdot q_{n, l}
\end{align}
\vspace{-15 px}
\begin{align}
\text{where}\ B_{n, k, l} = 2^{k+l}  \nonumber
\end{align}

\subsection{Compression Rate}

We use the compression rate (CR) to accurately reflect the amount of model size eliminated through pruning and quantization. The CR can be expressed in Equation~(\ref{eq:compression-rate}).
\begin{align}\label{eq:compression-rate}
\text{CR}_n = & \sum_i \bigl( N_{n,i}b_{max}-N_{n,i}b_n (1-p_{n,i})\bigr)/S
\end{align}
Here, $N_{n,i}$ is the number of weights in the $i$-th filter of the $n$-th layer of the model, and $S$ represents the total model size in terms of bits, which can be considered constant for a given target model.

\begin{figure}[t]
\centering
    \includegraphics[width=1\linewidth]{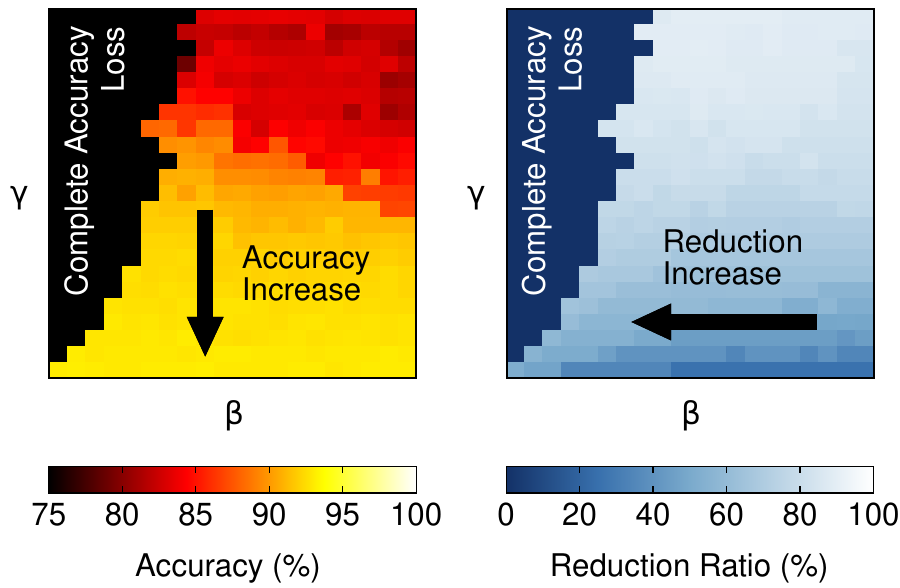}
    \caption{Heat maps of accuracy (left) and model reduction ratio (right) of ResNet-9 as distributions of hyperparameters $\beta$ (x-axis) and $\gamma$ (y-axis). Dark areas represent regions where accuracy is completely lost.}    \label{fig:hyperparameter-space}
\end{figure}

The CR can already be expressed in QUBO form, so we can directly obtain the third component, $R(\mathcal{P},\mathcal{Q})$, which is expressed in Equation~(\ref{eq:compression-layer}). The self-effect coefficients of qubits $p_{n,i}$ and $q_{n,k}$ are represented by $D_{n,i}$ and $E_{n,k}$, respectively. Additionally, the coefficient of interaction between two qubits, $p_{n,i}$ and $q_{n,k}$, is represented by $F_{n,i,k}$. These parameters can all be pre-calculated before coding for quantum optimization. Accordingly, the coupling map of qubits is shown in Table~\ref{table:qubo3}.
\begin{align} \label{eq:compression-layer}
    R ( \mathcal{P} , \mathcal{Q}) = & \sum_n \text{CR}_n \nonumber \\
     = &\sum_{n}\sum_{i}{D}_{n,i} \cdot \underline{p_{n,i}} + \sum_{n}\sum_{k}   {E}_{n,k} \cdot \underline{q_{n,k}} \nonumber \\
    -&\sum_{n}\sum_{i} \sum_{k} {F}_{n,i,k} \cdot \underline{p_{n,i}} \cdot \underline{q_{n,k}}
\end{align}
\vspace{-25 px}
\begin{align}
&\text{where}\nonumber \\
&{D}_{n,i} = \frac{N_{n,i}  b_{max}}{S}  ,\     {E}_{n,k} = \frac{\sum_i2^kN_{n,i}}{S} ,\  {F}_{n,i,k} = \frac{2^kN_{n,i}}{S}\nonumber  
\end{align}
The denominator $S$ in Equation (\ref{eq:compression-rate}) ensures that the model reduction ratio $R(\mathcal{P}, \mathcal{Q})$ is confined to the range $[0,1]$. Consequently, $R$ exhibits a consistent value range across different models, irrespective of their initial sizes.

\begin{algorithm}[t]
\caption{Hyperparameter Search Algorithm} \label{algo:hyperparameter}
\textbf{Input}:  \\
\hspace*{\algorithmicindent} $\mathbf{A}$, $\mathbf{B}$, $\mathbf{E}$, $\mathbf{F}$, $\mathbf{G}$: Hamiltonian coefficients. \\
\hspace*{\algorithmicindent} $a_{th}$: Accuracy threshold. \\
\textbf{Output}: \\
\hspace*{\algorithmicindent} $\gamma_{f},\: \beta_{f}$: Selected $\gamma$, $\beta$ values. \vspace{0.3em} \\
$\mathbf{procedure}$:\:\textsc{HyperparameterSearch}
\vspace{-1.2em}
\begin{algorithmic}[1] 
\STATE{Initialize $\beta \gets {\|\mathbf{A}\|_{1}}/{\|\mathbf{B}\|_{1}} $, $\gamma \gets random()$}.
\STATE{Get initial reduction rate: $\mathcal{P}, \mathcal{Q}, R \gets \mathbf{aqc}(\beta, \gamma)$}
\STATE{Get initial accuracy: $a \gets \mathbf{acc}(\mathcal{P}, \mathcal{Q})$}
\STATE{Initial checkpoint: $R_{f} \gets R$, \, $\beta_{f} \gets \beta$, \, $\gamma_{f} \gets \gamma$}
\FOR{$i = 1$ \TO $N_{iter}$} 
        \IF{$a \ge a_{th}$}
        \STATE{$\gamma_{l} \gets \gamma$}
        \STATE{$\gamma_{u} \gets \textsc{FindUpperGamma}(\gamma, \beta)$}
        \ELSE 
        \STATE{$\gamma_{l} \gets \textsc{FindLowerGamma}(\gamma, \beta)$}
        \STATE{$\gamma_{u} \gets \gamma$}
        \ENDIF
        \STATE{$\gamma \gets \textsc{BinSearchMax}(\gamma_{l}, \,\gamma_{u},\, a_{th})$}
        \STATE{$\beta \gets \textsc{BinSearchMin}(0,\, 2\beta, \,a_{th})$}
        \STATE{\textrm{\\}}
        \STATE{(Checkpoint solution if best compression)}
        \STATE{$\mathcal{P}, \mathcal{Q}, R \gets \mathbf{aqc}(\beta, \gamma)$}
        \STATE{$a \gets \mathbf{acc}(\mathcal{P}, \mathcal{Q})$}
        \STATE{$\textbf{if } R > R_{f} \textbf{ then } R_{f} \gets R, \, \beta_{f} \gets \beta, \, \gamma_{f} \gets \gamma$}
\ENDFOR
\end{algorithmic}
\end{algorithm}

\subsection{Combined QUBO Model} 

We combine the three energy components to yield the system Hamiltonian in QUBO form, shown in Equation~(\ref{eq:final-hamiltonian}):
\begin{align} \label{eq:final-hamiltonian}
    H = \mathbf{x}^T \mathbf{U}\mathbf{x},\ \ \ \textrm{where}\ \ \mathbf{x} = \text{vect}(\mathcal{P}, \mathcal{Q})
\end{align}
The operation $\text{vect}(\mathcal{P}, \mathcal{Q})$ strings together elements $p_{n,i} \in \mathcal{P}$ and $q_{n,k} \in \mathcal{Q}$ into a one-dimensional vector, such that each variable $p_{n,i}$ and $q_{n,k}$ is a unique qubit $x_{i}$. Accordingly, $\mathbf{U}$ is an upper-triangular matrix, in which each element is a linear combination of the three energy components. The coupling map of qubits for the formulation of $H$, and the details of each element in matrix $\mathbf{U}$ are provided in Table~\ref{table:qubo}. Given this coupling map, we can implement the QUBO function onto the quantum annealing machine, perform optimization and obtain the selected parameters for pruning and quantization on the model.

Implicitly, the hyperparameters $\beta$ and $\gamma$ are used to balance among the energy components in $H$. Hyper-parameter $\beta$ balances the weight of pruning loss and quantization loss. If $\beta$ is too large, we find that the system is averse to quantization and may over-prune the network without any quantization in the solution. Similarly, too small values of $\beta$ result in excessive quantization with minimal pruning. Instead, $\gamma$ functions largely as a design knob for us to trade off accuracy and model reduction ratio, as well as prevent over-compression phenomenon (complete loss of predictive power and can not be recovered).

\begin{algorithm} [t]
\caption{Upper/Lower Gamma Search Procedures} \label{algo:gamma}
\textbf{Input}: \\
\hspace*{\algorithmicindent} Current hyperparameter values $\gamma$, $\beta$. \\
\textbf{Output}: \\
\hspace*{\algorithmicindent} \textsc{FindUpperGamma}: $\gamma_{f}$ with invalid solution, $\gamma_{f} > \gamma$. \\ 
\hspace*{\algorithmicindent} \textsc{FindLowerGamma}: $\gamma_{f}$ with valid solution, $\gamma_{f} < \gamma$.  \vspace{0.3em} \\
$\mathbf{procedure}$:\:\textsc{FindUpperGamma} 
\begin{algorithmic}[1]
\STATE{Initialize $a \gets a_{th}$.}
\WHILE{$a \ge a_{th}$}
\STATE{$\gamma_{f} \gets 2 \gamma$}
\STATE{$\mathcal{P}, \mathcal{Q}, R \gets \mathbf{aqc}(\beta, \gamma)$}
\STATE{$a \gets \mathbf{acc}(\mathcal{P}, \mathcal{Q})$}
\ENDWHILE
\RETURN $\gamma_{f}$
\end{algorithmic}  \vspace{0.3em}
$\mathbf{procedure}$:\:\textsc{FindLowerGamma}
\begin{algorithmic}[1]
\STATE{Initialize $a \gets 0$.}
\WHILE{$a < a_{th}$}
\STATE{$\gamma_{f} \gets (\gamma / 2)$}
\STATE{$\mathcal{P}, \mathcal{Q}, R \gets \mathbf{aqc}(\beta, \gamma)$}
\STATE{$a \gets \mathbf{acc}(\mathcal{P}, \mathcal{Q})$}
\ENDWHILE
\RETURN $\gamma_{f}$
\end{algorithmic}
\end{algorithm}

\section{Hyper-Parameter Search} 
\label{section:hyperparameter-search}

From the QUBO Equation~(\ref{eq:hamiltonian-abstracted}), we note that the selection of appropriate hyperparameters is essential for AQC to yield a desirable compression solution. This fact is also observed in Figure~\ref{fig:hyperparameter-space} which shows a representative solution space (grid search on ResNet-9 at CIFAR-10 dataset) as hyperparameters $\gamma$ and $\beta$ are varied. These heat maps of accuracy and model reduction ratio are masked by an accuracy threshold, such that solutions with accuracy below the threshold are considered invalid and hidden. From the figure we observe that the process of determining suitable hyperparameters is itself a search problem and must be navigated efficiently by algorithms. Hence, we propose a simple but effective heuristic to quickly navigate the hyper-parameter space. Our observations on Figure~\ref{fig:hyperparameter-space} can be summarized as follows:
\vspace{-0.5em}
\begin{itemize}
    \setlength\itemsep{-0.2em}
    \item For each $\beta$ value (vertical column), there is a general increase in accuracy as $\gamma$ is reduced. This observation is intuitive as minimizing $\gamma$ reduces the impact of model reduction ratio on energy reduction during AQC.
    \item For each $\gamma$ value (horizontal row), there is an increase in model reduction ratio as $\beta$ is reduced, as the system becomes less averse to quantization.
    \item For each $\gamma$ value, there exists a minimum $\beta$ value above which the solution is valid. Below this $\beta$ value, the solution becomes over-compressed and loses inference ability completely. This region corresponds to the masked out (black/dark blue) region in Figure~\ref{fig:hyperparameter-space}.
\end{itemize}\vspace{-0.5em}

\begin{algorithm}[t]
\caption{Binary Search Procedure for Max/Min Value}  \label{algo:bin-search}
\textbf{Input}:  \\
\hspace*{\algorithmicindent} $x_{l},\: x_{u}$: Variable lower, upper bound. \\
\hspace*{\algorithmicindent} $a_{th}$: Accuracy threshold. \\
\textbf{Output}: \\
\hspace*{\algorithmicindent} $x_{f}$: Variable value after $N_{bin}$ iterations. \vspace{0.45em} \\ 
$\mathbf{procedure}$: \textsc{BinSearchMax}
\begin{algorithmic}[1]
\FOR{$i = 1$ \TO $N_{bin}$}
\STATE{$x_{t} \gets x_{l} + ({x_{u} - x_{l}})/{2}$}
\STATE{Evaluate accuracy $a$ using $\mathbf{aqc}(\cdot)$ and $\mathbf{acc}(\cdot)$. }
\STATE{$\textbf{if} \; a > a_{th} \;\textbf{then} \; x_{l} \gets x_{t}$}
\STATE{$\textbf{else} \; x_{u} \gets x_{t}$}
\ENDFOR
\RETURN $x_{l}$
\end{algorithmic} \vspace{0.3em}
$\mathbf{procedure}$: \textsc{BinSearchMin}
\begin{algorithmic}[1]
\FOR{$i = 1$ \TO $N_{bin}$}
\STATE{$x_{t} \gets x_{l} + ({x_{u} - x_{l}})/{2}$}
\STATE{Evaluate accuracy $a$ using $\mathbf{aqc}(\cdot)$ and $\mathbf{acc}(\cdot)$. }
\STATE{$\textbf{if} \; a > a_{th} \;\textbf{then} \; x_{u} \gets x_{t}$}
\STATE{$\textbf{else} \; x_{l} \gets x_{t}$}
\ENDFOR
\RETURN $x_{u}$
\end{algorithmic}
\end{algorithm} 

On the basis of these observations, we employ an iterative search to find suitable values of $\beta$ and $\gamma$. This iterative search is shown in Algorithm \ref{algo:hyperparameter}. The function call $\mathbf{aqc}(\cdot)$ performs AQC on the Hamiltonian generated using $\beta$ and $\gamma$, returning solution variables $\mathcal{P}$, $\mathcal{Q}$, and model reduction ratio of the solution $R$. Meanwhile, function call $\mathbf{acc}(\cdot)$ generates the compressed model using $\mathcal{P, Q}$, fine-tunes the model for one epoch, and then evaluates the compressed model on the test dataset to yield the resulting accuracy $a$.

Algorithms \ref{algo:gamma} and \ref{algo:bin-search} further detail procedures used in the main algorithm. The procedures $\textsc{FindUpperGamma}$ and $\textsc{FindLowerGamma}$ in Algorithm \ref{algo:gamma} yield appropriate search bounds of hyperparameter $\gamma$. The procudures  $\textsc{BinSearchMax}$ and $\textsc{BinSearchMin}$ in Algorithm \ref{algo:bin-search} presents the binary search processes to maximize or minimize the solution variable respectively.

Refer to Algorithm~\ref{algo:hyperparameter}, we begin with an initial $\beta$ value of ${\|{\mathbf{A}}\|_{1}}/{\|\mathbf{B}\|_{1}}$ and an arbitrary $\gamma$ value (line 1). As coefficients ${A}$ and ${B}$ approximate the severity of pruning loss and quantization error respectively, we hypothesize that the $l1$-norm of these coefficients across the model should be similar in magnitude to balance between pruning and quantization. From our experiments, we find that this initial $\beta$ hypothesis is generally near-optimal, providing a useful starting point for the search.

\begin{figure*}[t]
\centering
    \includegraphics[width=1\linewidth]{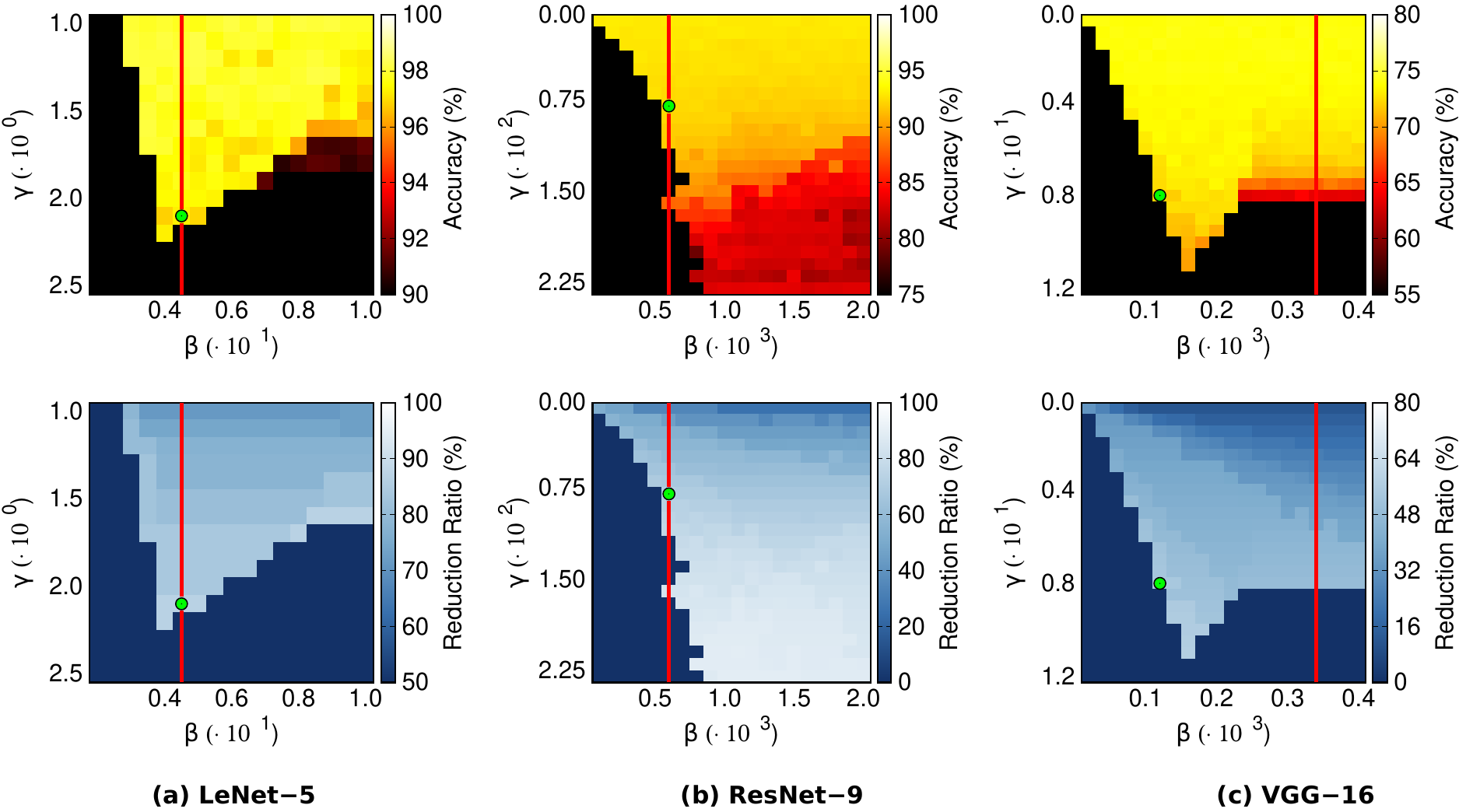}
    \caption{Heat maps of accuracy (red, top row) and model reduction ratio (blue, bottom row) of (a) LeNet-5 for MNIST, (b) ResNet-9 for CIFAR-10, and (c) VGG-16 for CIFAR-100. The objective is to select a point that is brightly-colored in both accuracy (top) and model reduction ratio (bottom) heat maps. The red vertical line on each plot indicates the proposed $\beta$ initialization as the starting point for our proposed hyper-parameter search algorithm. The green dot marks the final solution found.}
    \label{fig:plots}
\end{figure*}

Using these initial values as starting hyperparameters, we iteratively perform the following:
\vspace{-0.5em}
\begin{enumerate}

\item Fixing $\beta$, determine an appropriate range of $\gamma$ values in which to search (lines 6-12). Specifically, we aim to search in the range of $[\gamma_{l}, \, \gamma_{u}]$, for which $\gamma_{l}$ has high accuracy above threshold $a_{th}$ while $\gamma_{u}$ has accuracy below threshold $a_{th}$
.
\item Perform binary search for the maximum $\gamma$ in the range $[\gamma_{l}, \, \gamma_{u}]$ that fulfills the accuracy threshold (line 13). Let this value be $\gamma_{n}$

\item Fixing $\gamma$, perform binary search for minimum $\beta$ in the range $(0, 2\beta]$ that fulfills the accuracy threshold (line 14). We increase the upper bound to $2\beta$ in order to explore solutions with less quantization than the current $\beta$ value. Let the $\beta$ found through binary search be $\beta_{n}$.

\item (If necessary) Repeat steps 1 to 3 for $N_{iter}$ times, fixing $\beta = \beta_{n}$.
\end{enumerate}

In each step, we perform quantum optimization using $\mathbf{aqc(\cdot)}$ to yield a pruning-quantization solution (line 17), and then fine-tune the model for one epoch and assess solution accuracy (line 18). Three user-defined search configurations include the accuracy threshold $a_{th}$ for valid solutions, the number of binary search iterations $N_{bin}$, and the number of overall exploration iterations $N_{iter}$ for repeat steps $1$ and $2$. While the proposed procedure is iterative, we find that our initialization of $\beta = {\|\mathbf{A}\|_{1}}/{\|\mathbf{B}\|_{1}}$ provides a good starting point and often yields the near-optimal solution within the first search iteration.

\renewcommand{\arraystretch}{1.45}
\begin{table*}[h]
\caption{Accuracy and model reduction ratio of solutions from Quantum/Digital annealing and traditional solutions. 'Problem size' denotes the number of variables (qubits). 'Weight pruning'~\cite{yang2020automatic} is more fine-grained and can typically achieve a higher reduction ratio but results in irregular computation dataflow during implementation.} \label{table:compression-results}
\vspace{0em}
\begin{center}
\resizebox{1.00\linewidth}{!}{\begin{tabular}{l|c|c|c|c|c|c|c|c}
\hline
\hline
\multirow{2}{*}{\textbf{Model}}   & \multirow{2}{*}{\textbf{Dataset}} & \multirow{2}{*}{\textbf{Method}} & \textbf{Pruning}   &\multirow{2}{*}{\textbf{Quantized}} & \textbf{Problem} & \multirow{2}{*}{\textbf{Accuracy}} & \multirow{2}{*}{\textbf{Drop}} & \textbf{Reduced} \vspace{-4.5pt}\\
{} & {} &   & \textbf{Granularity} & & \textbf{Size} &  &  & \textbf{Ratio$^3$}  \\
\hline
\multirow{4}{*}{LeNet-5}   & \multirow{4}{*}{MNIST}   &\textbf{Ours$^1$}                 & Filters   &\checkmark & 28    & 98.17\% &-0.38\%   & 96.5\%  \\
\cline{3-9}
                           &                          &\textbf{Ours$^1$}                 & Channels  &\checkmark & 108   & 97.98\% &-0.57\%   & 96.8\%  \\
\cline{3-9}
                           &                          &\textbf{Ours$^2$}                 & Filters   &\checkmark & 28   & 98.20\% &-0.35\%    & 95.6\%  \\
\cline{3-9}
                           &                          &\textbf{Ours$^2$}                 & Channels  &\checkmark & 108   & 98.22\% &-0.33\%   & 96.7\%  \\  
\hline
\multirow{2}{*}{GTSR-CNN}  & \multirow{2}{*}{GTSRB}   &\textbf{Ours$^1$}                 & Filters   &\checkmark & 233   & 98.00\% &-0.57\%   & 91.8\%  \\
\cline{3-9}
                            &                         &\textbf{Ours$^2$}                 & Filters   &\checkmark & 233   & 98.10\% &-0.47\%   & 90.2\%  \\
\hline
{ResNet-9}                  & {CIFAR-10}              &\textbf{Ours$^2$}                 & Filters   &\checkmark & 2264  & 92.40\% &-0.57\%   & 93.1\%  \\
\hline
\multirow{3}{*}{VGG-16}    & \multirow{3}{*}{CIFAR-100}&\textbf{Ours$^2$}                & Filters   &\checkmark & 4263  & 73.15\% &-0.89\%   & 88.4\%  \\
\cline{3-9}
                           &                          &ANNC \cite{yang2020automatic}     & Weights   &\checkmark & N.A.  & 71.59\% &-2.45\%   & 89.7\%  \\
\cline{3-9}
                           &                          &CAC \cite{chen2020dynamical}      & Filters   &$\times$   & 4224  & 72.28\% &-1.76\%   & 55.6\%  \\
\hline
\hline
\multicolumn{9}{l}{$^1$ D-wave Advantage; \ \ \ \ \ \ $^2$ Fujitsu DA; \ \ \ \ \ \ $^3$ Compared with FP32 models}
\end{tabular}}
\end{center}
\end{table*}

\section{Experiments} \label{section-Results}

In the experiment, we include the {D-Wave Advantage system~\cite{dwave_arch}, which is based on a lattice of interconnected qubits. Each qubit is made of a superconducting loop. For QPU, we set the ``number of reads'' to 32 for all neural networks, translating to 32 samples for every optimization task. We use 20 $\mu s$ for all tasks' annealing time and default annealing schedule. We included the Fujitsu Digital Annealer (DA)~\cite{tsukamoto2017accelerator} to evaluate problem sizes too large for the D-Wave Advantage of today. The enhancements include but are limited to parallel-trial, dynamic offset, and Parallel Tempering (PT) with Isoenergetic Cluster Moves (ICM)~\cite{zhu2020borealis,aramon2019physics}. For DA, we set the ``number of replicas'' to $32$ for all neural networks, which produces $32$ samples for every optimization task. We use $10^8$ number of sweeps (or iterations in Fujitsu's term)  to ensure the system rests well near the ground state of the energy landscape of an optimization problem.

\subsection{Models and Datasets}

The different annealing technologies impose constraints on the neural network architectures due to the limited number of nodes and their connectivity. Nonetheless, we demonstrate the performance of compression by AQC across four different datasets: MNIST, CIFAR-10, CIFAR-100, and the application-oriented GTSRB. For this task of the German Traffic Sign Recognition Benchmark (GTSRB)~\cite{stallkamp2011german}, we use a custom small CNN (henceforth GTSR-CNN) consisting of three convolution layers (with 32, 64, and 128 output channels, respectively) and two linear layers. Due to constraints in qubit count and connectivity in commercially available annealing devices~\cite{dwave_arch,tsukamoto2017accelerator}, we selected models with sufficiently small problem size to be implemented, which are LeNet-5, ResNet-9, and VGG-16. In our context, problem size denotes the \emph{number of variables (qubits)} required to solve the (filter) pruning-quantization of the model, and not the size of model parameters. Our model reduction rate is calculated assuming that the original weights are in INT8 format. However, in subsection~\ref{ss:sota}, for comparison with other state-of-the-art work, the original weights are assumed to be in FP32 format, representing the model size after pre-training.

Our models are pre-trained with 8-bit weights to reach state-of-the-art accuracy equivalent to floating-point weights. Using our search algorithm, we then determine appropriate $\gamma$ and $\beta$. During hyper-parameter search, the accuracy threshold for valid solutions ($a_{th}$) is set to 2\% below original accuracy for LeNet-5 and GTSR-CNN, 4\% loss for ResNet-9, and 10\% for VGG-16. The lower accuracy thresholds allow for higher-compression solutions to be accepted, as the accuracy can be restored during fine-tuning. Iteration configurations $N_{bin}$ and $N_{iter}$ are both set to $5$, although the final solutions were found within the first 1-2 iterations. We provide optimization results from two devices. LeNet-5 and GTSR-CNN are executed on both Fujitsu's and D-Wave's devices, while the larger models are run solely on Fujitsu's devices due to the capacity limitations of D-Wave's device. Using the final solution, we fine-tune the compressed model until convergence. 

\subsection{Hyper-Parameter Search}

We assess our proposed search algorithm on the hyper-parameter solution space of LeNet-5, ResNet-9, and VGG-16. First, we perform a coarse-grained grid search over a suitable range of hyperparameters $\beta$ and $\gamma$ to determine the accuracy and compression performance landscape. Several heat maps from our grid search are shown in Figure~\ref{fig:plots}. The top row of plots (red heat maps) indicates the accuracy of compressed models found using the corresponding $\beta$ and $\gamma$ values. The entire regions on the heat map have been scaled appropriately to visualize the accuracy distribution better. The bottom row of plots (blue heat maps) shows the corresponding model reduction ratio. In both heat maps, the masked (black/dark blue) regions indicate hyper-parameter values upon which the model loses predictive capabilities.

As discussed in Section \ref{section:hyperparameter-search}, the plots illustrate the accuracy-compression trade-off provided by tuning hyperparameters $\beta$ and $\gamma$. The objective is to find a solution point $(\beta, \gamma)$ with a high value (bright color) in both accuracy and model reduction ratio. In Figure~\ref{fig:plots}, we indicate with vertical lines the proposed initial $\beta$ value from our heuristic of balancing the $l1$-norm pruning and quantization energy coefficients. We also denote the final compression solution found using green dots. From the plots, we observe that our algorithm can find solutions near the optimal region of the accuracy-compression trade-off within the given grids. Moreover, our preliminary guess of $\beta = {\|\mathbf{A}\|_{1}}/{\|\mathbf{B}\|_{1}}$ provides an appropriate starting point, even being situated near the optimal trade-off region in the cases of LeNet-5 and ResNet-9. However, for VGG-16, our algorithm can step towards a better solution away from the initial guess of $\beta$ as well.

\subsection{Compression Performance by AQC}
\label{ss:sota}
\begin{figure*}[t]
\centering
    \includegraphics[width=1\linewidth]{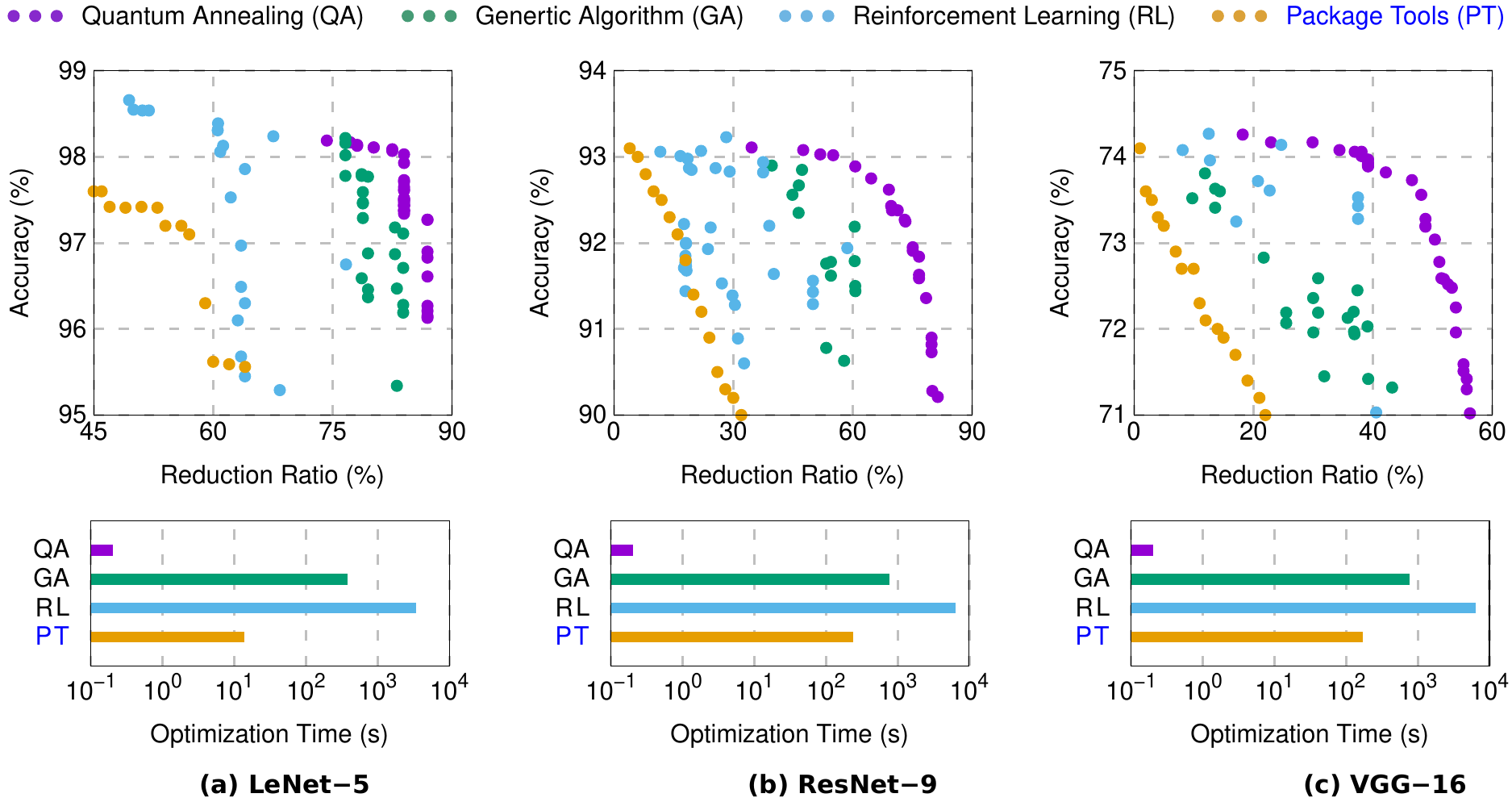}
    \caption{Comparison of quantum annealing (QA), genetic algorithms (GA), reinforcement learning (RL), and {machine learning package tools (PT)} regarding accuracy, model reduction ratio, and optimization time. We implemented the GA algorithms (NSGA-II) using code from Reference~\cite{wang2021evolutionary} and RL algorithms (SAC) using code from Reference~\cite{wang2022edcompress}, tuning hyperparameters until the optimization converged. {For the PT method, we utilize quantization aware training (QAT)~\cite{pytorch_q} and L$_n$-structured pruning~\cite{pytorch_p} algorithms, both integrated into the PyTorch framework.}}
    \label{fig:overcompare}
\end{figure*}

Table~\ref{table:compression-results} presents the accuracy and model reduction ratios of solutions found by both platforms. We perform quantum compression using our formulation on the D-Wave Advantage quantum computer and the Fujitsu annealer. We observe that our AQC formulation can provide high compression rates on these models while maintaining minimal accuracy loss of less than 1\% from the full-precision model used as the starting point. As most larger models were limited to filter pruning due to AQC capacity constraints, we investigated channel-wise pruning of LeNet-5 (i.e., channels of filters pruned individually). LeNet-5 was chosen for channel-wise pruning as the problem size is small enough to run on the current D-Wave Advantage system. Our formulation could also perform compression at channel-wise granularity, demonstrating the potential to use quantum optimization for even finer-grained pruning-quantization. With our algorithm, we can easily perform compression on large models or at per-weight granularity once quantum annealing machines with extensive qubit scale and connectivity become commercially feasible and available.

Lastly, we compare our compression performance to two recent works. It is challenging to make a direct comparison due to differences in compression methods, supported models, and compression granularity. We compare our work to ANNC~\cite{yang2020automatic} which has the similar objective of finding pruning-quantization ratios in an automated manner, although they perform pruning at the per-weight granularity. We run the original source code on VGG-16/CIFAR-100, tuning the target sparsity and bit budget to achieve similar accuracy to our work. However, for ANNC on VGG-16, we could not find a configuration to achieve accuracy greater than 72\% at similar model reduction ratio in our experiments. Nonetheless, our work could generate solutions with magnitudes of accuracy and compression close to that of ANNC despite being limited to coarse-grained filter pruning. We also compare our work to CAC~\cite{chen2020dynamical}, which performs filter pruning but without quantization. The accuracy and pruning rate are obtained from their published results. With different objectives (pruning only vs pruning-quantization), CAC achieves a much lower model reduction ratio due to its full precision weights.

\subsection{Compared with Traditional Approach}

To further explore the effectiveness and time efficiency of quantum annealing in model compression, we implemented two classical approaches to solve large space optimization problems: genetic algorithms (GA) and reinforcement learning algorithms (RL). We applied quantum annealing (QA), GA (NSGA-II)~\cite{wang2021evolutionary}, and RL (SAC)~\cite{wang2022edcompress} to the same model under identical conditions and compared the optimization results. The comparison results are shown in Figure~\ref{fig:overcompare}. In the first row, we plot the results for two objectives of the optimization: accuracy, which aims for higher values, and model reduction ratio, where larger is considered better. In the second row, we compare the processing times. For the GA and RL algorithms, which base their optimization results on intermediate outcomes, we derive actual accuracy from the model itself rather than evaluating WM and RMSE, as is done in quantum annealing.

Results in Figure~\ref{fig:overcompare} clearly demonstrate the advantages of quantum annealing compared to traditional approaches. Although in some cases, such as with LeNet-5 and VGG-16, the RL algorithm can achieve higher accuracy, it does not offer an efficient compression rate compared to quantum annealing. For these traditional methods, it is more difficult to approach the optimal point where the model reduction ratio can be maximized while still maintaining model accuracy. In terms of processing time, our quantum annealing approach demonstrates a significant speedup, attributed to the physical characteristics of the computation mechanism. Within a finite optimization time, the quantum-based approach can fully explore the optimal space, avoiding local optima, and outperforms those traditional optimization methods, even though they are specifically designed for large search spaces.

{Compared with Classical Package Tools}}
{
We also incorporate package tools (PT) into the comparison. Specifically, we utilize quantization-aware training (QAT)~\cite{pytorch_q} and structured pruning (L$_n$-structured)~\cite{pytorch_p} algorithms integrated into the PyTorch platform. We conduct multiple independent experiments; in each experiment, the model is quantized into the INT8 data format, and the same pruning amount ratio is applied to all layers in the model. We test different pruning amounts in different experiments. Each experiment corresponds to a dot in the figures. At the beginning of each experiment, we reload the well-trained full-size model under FP32 data format, apply integrated quantization and pruning functions, then fine-tune the model until it reaches its maximum accuracy. 
}
{
From Figure~\ref{fig:overcompare}, we can see that the optimization results under the platform tools have weaker performance compared to the other three methods. This is because this method lacks variation in compression parameters (quantization bits and pruning amount) across different layers of the model. Allowing this diversity would significantly enlarge the search space, making manual tuning infeasible. On the other hand, the genetic algorithms method, reinforcement learning method, and our quantum annealing-based method are all automated tuning algorithms capable of optimizing compression parameters from this vast search space. This is why they show better results.
}

\section{The Readiness of Quantum Optimization}

Through our efforts in problem reformulation and hyper-parameter search, we find that AQC can provide effective solutions for neural network compression. These results demonstrate the potential for using quantum optimization for complex deep learning applications when quantum computing reaches greater maturity. However, there remain several constraints that limit the capabilities and, thus, adoption of quantum optimization for these tasks to date. In this section, we evaluate the readiness of quantum optimization from our observations in model compression, and highlight future advancements necessary for quantum optimization to be translated to more problem settings and applications. 

\subsection{Number of Qubits}

The number of qubits available on quantum devices today naturally limit the problem sizes that can be optimized. In our experiments in model compression, we find that AQC can already yield promising solutions for small-scale models with few layers, such as those used for symbol recognition tasks (MNIST, GTSRB). The size of densely-connected optimization problems that we could implement on the D-Wave Advantage quantum annealer were at several hundred solution variables. However, problems of small scales can be partially handled by classical computing platforms today. Further technological advancement to increase qubit numbers will be essential to realize the promise of solving massive, complex problems by quantum optimization.

\subsection{Qubit Connectivity}

In current quantum hardware, qubits are not fully connected but instead follow a fixed topology, thus limiting inter-qubit interactions. As a result, current quantum optimization is well-positioned to solve problems with limited connectivity between solution variables. For example, the D-Wave Advantage with over 5000 qubits could solve problems with several thousand solution variables if the QUBO is sparse and topologically compatible. However, many real-world optimization problems (including model compression) are densely-connected with dependence between many solution variables. The embedding of dense variables to physical qubits typically results in a great increase in qubits and couplers required. In our experiments, the D-Wave Advantage did not have sufficient capacity to process ResNet-9 (2264 logical qubits) and VGG-16 (4263 logical qubits) after the process of hardware mapping.

\subsection{Accessibility and Ease of Programming}

Current quantum devices are at an early stage of development. As such, they are currently deployed as shared cloud-based platforms, which limits the availability and degree of custom configuration provided to the general user. In our experiments, while the true optimization time was short (in the order of milliseconds) - a defining advantage of quantum optimization, a much longer time was required for queuing, scheduling and data transfers to the cloud platform even for small problems. While platform development has progressed significantly in recent years, further improvements to accessibility, reconfigurability, and the programming interface would help to increase adoption of quantum optimization in more problem settings. It remains uncertain whether future quantum technologies will support highly-configurable on-premise devices.

\section{Conclusion}

In this work, we investigate the potential of quantum optimization for joint pruning-quantization of neural network models. As model compression is not naturally compatible with AQC, we reformulate the problem as a QUBO, allowing the accuracy-compression trade-off to be efficiently navigated using quantum processes. We find that quantum optimization can potentially provide quality compression solutions for practical small-scale neural networks. However, our experiments remain limited by current constraints of annealing technologies, including qubit count and connectivity. As quantum technology matures, we hope that our introduction of AQC for neural network design can spark new works to further advance quantum optimization towards supporting larger, more complex deep learning problems in time to come.

\bibliographystyle{cas-model2-names}
\bibliography{Ref2}

\end{document}